\documentclass[superscriptaddress,preprint, showpacs,amsmath,amssymb, aps,prb,longbibliography,lengthcheck]{revtex4-1}

\usepackage{color}
\usepackage{graphicx}
\usepackage{dcolumn}
\usepackage{bm}
\usepackage{hyperref}
\usepackage[normalem]{ulem} 

\begin{document}

\title{Scaling Study and Thermodynamic Properties of the cubic Helimagnet FeGe}

\author{H. Wilhelm}
\affiliation{Diamond Light Source Ltd., Chilton, Didcot, Oxfordshire, OX11 0DE,
United Kingdom}

\author{A. O. Leonov}
\affiliation{Zernike Institute for Advanced Materials, University of Groningen, Nijenborgh 4, 9747 AG Groningen, The Netherlands}
\affiliation{Center for Chiral Science, Hiroshima University, Higashi-Hiroshima, Hiroshima 739-8526, Japan}

\author{U. K. R\"{o}{\ss}ler}
\affiliation{IFW Dresden, P.O. Box 270116, 01171 Dresden, Germany}

\author{P. Burger}
\affiliation{Institute for Solid State Physics, Karlsruhe Institute of
Technology, P.O. Box 3640, 76021 Karlsruhe, Germany}

\author{F. Hardy}
\affiliation{Institute for Solid State Physics, Karlsruhe Institute of
Technology, P.O. Box 3640, 76021 Karlsruhe, Germany}

\author{C. Meingast}
\affiliation{Institute for Solid State Physics, Karlsruhe Institute of
Technology, P.O. Box 3640, 76021 Karlsruhe, Germany}

\author{M. E. Gruner}
\affiliation{Faculty of Physics and Center for Nanointegration Duisburg-Essen (CENIDE), University of Duisburg-Essen, 47048 Duisburg, Germany}

\author{W. Schnelle}
\affiliation{Max Planck Institute for Chemical Physics of Solids,
N\"{o}thnitzer-Str. 40, 01187 Dresden, Germany}

\author{M. Schmidt}
\affiliation{Max Planck Institute for Chemical Physics of Solids,
N\"{o}thnitzer-Str. 40, 01187 Dresden, Germany}

\author{M. Baenitz}
\affiliation{Max Planck Institute for Chemical Physics of Solids,
N\"{o}thnitzer-Str. 40, 01187 Dresden, Germany}

\begin{abstract}
The critical behavior of the cubic helimagnet FeGe was obtained from isothermal magnetization data in very close vicinity of the ordering temperature.
A thorough and consistent scaling analysis of these data revealed the critical exponents $\beta=0.368$, $\gamma=1.382$, and $\delta=4.787$.
The anomaly in the specific heat associated with the magnetic ordering can be well described by the critical exponent $\alpha=-0.133$.
The values of these exponents corroborate that the magnetic phase transition in FeGe belongs to the isotropic 3D-Heisenberg universality class.
The specific heat data are well described by \emph{ab initio} phonon calculations and confirm the localized character of the magnetic moments.

\end{abstract}



\maketitle

\section{Introduction}
The magnetic properties of the chiral helimagnet FeGe\cite{Lundgren70,Bak80,Ericsson81,Lebech89,Plumer90} are of renewed interest after the prediction of Skyrmion ground states in cubic helimagnetic metals\cite{Roessler06}.
These vortex-like magnetic structures were predicted to exist at temperatures ($T$) just below the critical temperature ($T_c$) and for weak magnetic fields ($H$) in a narrow pocket of the ($H$,$T$) phase diagram.
Fingerprints of this new magnetic state have been observed in thin films of FeGe by Lorentz transmission-electron microscopy\cite{Yu10,Zhao16} and in bulk samples by small-angle neutron scattering\cite{Moskvin13} as a characteristic distribution of the lateral magnetization and a hexagonal Bragg-spot pattern, respectively.
The occurrence of a Skyrmion phase and the helical magnetism in zero field are the consequence of the Dzyaloshinskii-Moryia interaction (DM).
Its strength together with the exchange interaction governs the pitch of the spin spiral, the diameter of a single Skyrmion, and the saturation field ($H_{c2}$) required to fully align the magnetic moments along the external field\cite{Wilhelm11,Wilhelm12}.
Like in other cubic chiral magnets crystallizing in non-centrosymmetric crystal structures, a confined precursor region exists in a narrow temperature interval between the ordered and paramagnetic (PM) state.
Here the longitudinal magnetization strongly varies and short-ranged chiral spin correlations prevail.
This region can be identified as an inhomogeneous chiral-spin (ICS) state.
In FeGe it extends from $T_c=278.2(3)$~K up to $T_0\simeq 280$~K\cite{Wilhelm11,Wilhelm12,Cevey13,Barla15}.

Despite the interest in the field-induced complex magnetic structures of FeGe, several basic thermodynamic properties and the transition to the helical state in zero field have not been studied in great detail.
In this context the investigation of the critical behavior of FeGe and related cubic helimagnets is of prime  interest as it yields important microscopic information about the underlying magnetic interactions.

Several studies were devoted to determine the critical exponents of some chiral cubic magnets.
Nonetheless, the reported values and universality classes across these compounds are controversial and contradictory.
Based on isothermal magnetization data, it was suggested that MnSi belongs to the tri-critical mean-field universality class\cite{Zhang15} whereas FeGe was proposed to change the universality class across the phase transition from 3D-Heisenberg (above $T_c$) to 3D-Ising or 3D-XY (below $T_c$)\cite{Zhang16}.
However, a study on Fe$_{0.8}$Co$_{0.2}$Si using isothermal magnetization and magneto-transport reported a consistent set of critical exponents that fall in the 3D-Heisenberg universality class\cite{Jiang10}.
Results of a scaling analysis based on temperature-dependent ac-susceptibility data on Cu$_2$OSeO$_3$ led to the conclusion that this insulating chiral magnet also belongs to the 3D-Heisenberg universality class\cite{Zivkovic14}.

In particular the inconsistent findings reported for FeGe\cite{Zhang16} are puzzling as it can be regarded as a strong band-ferromagnet with a stable magnetic moment ($\mu_{\mathrm{Fe}}=1\mu_{\mathrm{B}}$), a relatively high ordering temperature, and negligible spin fluctuations\cite{Lundgren70,Ericsson81,Barla15,Yamada03,Neef09}.
In this respect FeGe is similar to a classical band-ferromagnet (FM), like Ni\cite{Seeger95}, but with the peculiarity that the transition from the PM phase to helical ordering is obscured by the ICS state.
Therefore, a consistent scaling analysis of the critical behavior should be performed in fields strong enough to overcome the DM interaction and to exclude any influence of the helical, conical, and skyrmion states.
Thus, the criticality of the unperturbed magnetic spin configuration should be studied at fields well above $H_{c2}(T=0)=3.6$~kOe\cite{Wilhelm12} and for temperatures within a distance $\epsilon=|1-T/T_c|\lesssim 10^{-2}$ to $T_c$ as for a classical FM to PM transition at elevated temperatures\cite{Seeger95}.

In the following we present a consistent scaling analysis of FeGe to determine its universality class.
This investigation is based on isothermal magnetization data, $M(H)$, very close to $T_c$ ($|\epsilon| < 0.013$) and specific heat data in zero field.
Furthermore, the lattice contribution to the specific heat and thermal expansion is compared with \emph{ab initio} phonon calculations in the quasi-harmonic approximation\cite{Fultz10}.

\section{Methods}
\subsection{Experimental Details}
In this work high-quality single crystals grown by chemical vapor transport\cite{Wilhelm07} have been used.
The zero-field cooled isothermal magnetization data were measured on the same crystal that was used in the study reported earlier\cite{Wilhelm11}.
Its mass was 0.995~mg and it was carefully aligned with its [100] direction along the external magnetic field.
Isothermal $M(H)$ data were recorded with a commercial SQUID magnetometer (MPMS Quantum Design) after zero field cooling from 320~K.
After each data set had been completed the field was swept to zero in an oscillating fashion and then the sample was heated to 320~K where it was kept for five minutes before the next cycle was started.
In very close vicinity of $T_c$, i.e., 275~K~$\leq T \leq 283$~K, the isotherms were recorded as close as $\Delta T=0.2~\mathrm{K}$ whereas further away from $T_c$ they were at least $\Delta T\geq 1~\mathrm{K}$ apart.
The field was incremented initially by $\Delta H=500~\mathrm{\mathrm{Oe}}$ and beyond $H>10~\mathrm{kOe}$ in steps of $\Delta H=2~\mathrm{kOe}$.
The crystal had a shape of an irregular but almost spherical polyhedron and therefore the demagnetization factor $N=1/3$ was used to determine the internal magnetic field according to $H= H_{\mathrm{appl}}-N\times \rho M$, with $M$ the mass magnetization (in units of emu/g) and $\rho = 8.22$~g/cm$^3$, the mass density of FeGe.

The specific heat in zero field was measured on three single crystals ($m_{\mathrm{tot}}=6.07$~mg) in a PPMS Quantum Design device using the heat-pulse method.
One single crystal, which had an arbitrary orientation with respect to the field, was used for the specific heat measurements in external magnetic fields.
The thermal dilation measurements were carried out with a home-made capacitive dilatometer using the parallel-plate capacitance method\cite{Pott83,Meingast90,Meingast01}.
The measurements were done along the longest direction ($\approx 0.8$~mm) of two opposite parallel surfaces of a polyhedral shaped single crystal.
The temperature was increased at a rate of 15~mK/s and averages were made every 100~mK.

\subsection{First-Principal Calculations}
\label{sec:firstprincipalcalculations}
The phonon density of states was calculated
numerically within the direct method\cite{Kresse95,Gonze97,Parlinski97}.
Here, the phonon frequencies were calculated from restoring forces
generated by a distortion of the ideal crystal lattice
due to small atomic displacements.
A Fourier transform of the force-constant matrix yields the dynamical
matrix and its diagonalization gives the wave-vector dependent phonon
frequencies.
The forces were obtained from first-principles supercell calculations within the framework of density functional theory (DFT) with the Vienna \emph{ab initio} Simulation Package\cite{cn:VASP1,cn:VASP2}.
A scalar relativistic description was used in connection with the Generalized Gradient Approximation (GGA) by Perdew, Burke and Ernzerhof\cite{cn:Perdew96}.
The cutoff for the plane wave basis functions was chosen as $E_{\mathrm{cut}}=400\,$eV.
The semi-core $3p$ for Fe and $3d$ for Ge, i.e., $3p^63d^74s^1$ and $3d^{10}4s^24p^1$, respectively, were considered explicitly in the potentials.
The calculations were done with a supercell of 3$\times$3$\times$3 primitive cells, containing 216 atoms in total.
For sufficient accuracy of the forces a $k$-mesh of 4$\times$4$\times$4 points in the full Brillouin zone was employed in connection with the Methfessel-Paxton finite temperature integration scheme (smearing parameter $\sigma$$\,=\,$0.1\,eV).
The forces were calculated from four displacements of $0.04$\,\AA\,in size, each.
In order to minimize errors from noise and anharmonicities, displacements in opposite directions were considered.

The PHON code by D. Alf\`{e}\cite{cn:Alfe09} was employed in combination with the PHONON code by K. Parlinsky\cite{Parlinski_M} to generate the displacements, compute the phonon dispersion relations and calculate the  free energy, specific heat, and entropy at a fixed volume.
To obtain thermodynamic averages in the quasiharmonic approximation\cite{Fultz10,Grimvall86} phonons were calculated  for five different volumes of the primitive cell.
The temperature and volume dependent free energy was interpolated using a
third order spline scheme.
Minimization with respect to the volume at each constant temperature yields access to the equilibrium volume, entropy, and the specific heat at zero constant pressure.

\section{Results and discussion}
\subsection{Magnetization}\label{sec:arrott}
%
\begin{figure}
\includegraphics[width=8.5cm]{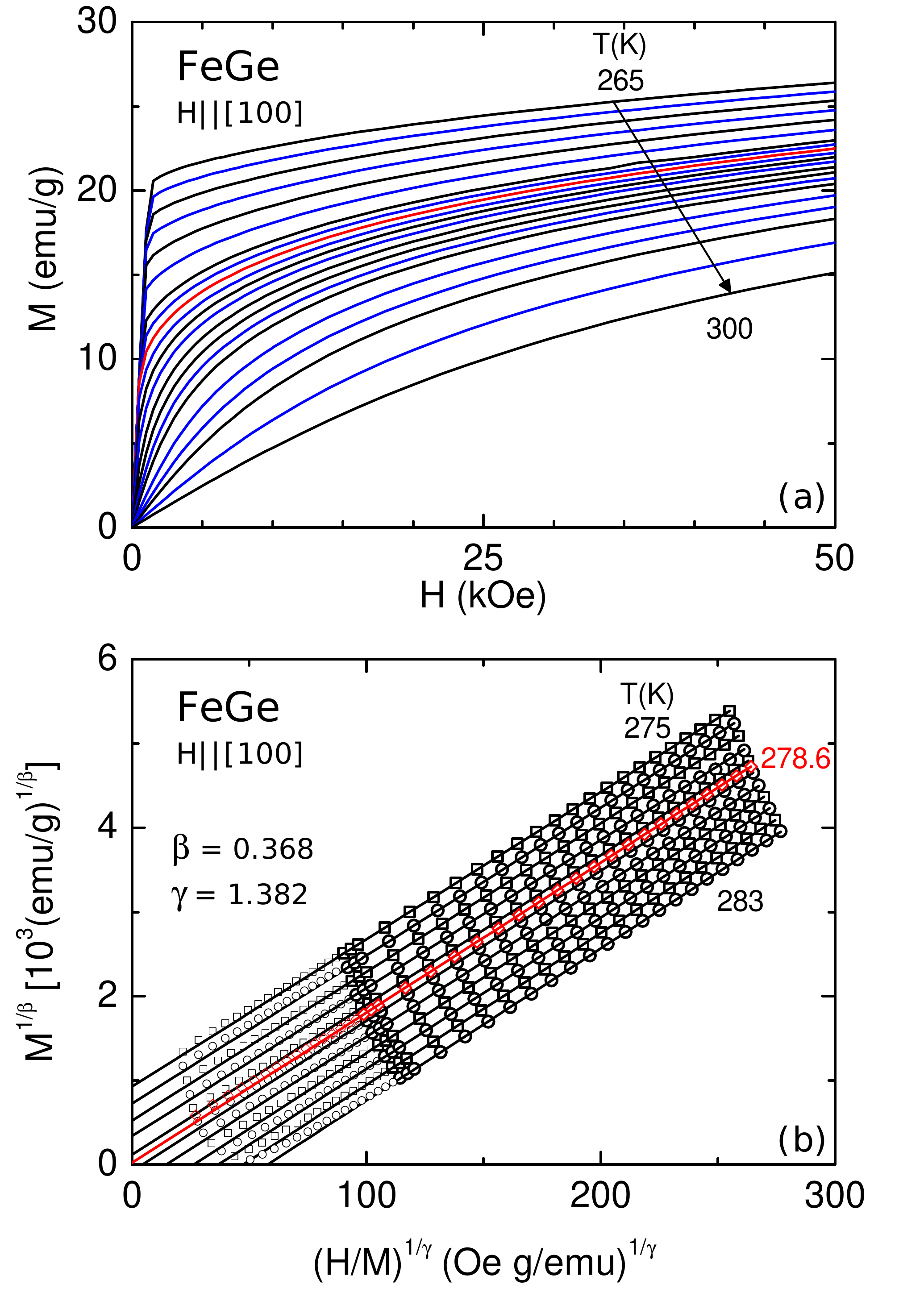}
\caption{(a) Isothermal magnetization ($M$ vs.\,$H$) of FeGe at selected temperatures\cite{Tsteps} for magnetic fields $H\parallel[100]$.
(b) Re-scaled isotherms using the critical exponents $\beta=0.368$ and $\gamma=1.382$.
Only a subset of the measured data is shown for clarity\cite{Tsteps}.
The lines are linear fits to the data (large symbols) for fields above $H_{\mathrm{cutoff}}=9$~kOe.
In both panels the critical isotherm is highlighted in red and in (b) the applied magnetic field was corrected for demagnetization effects.
\label{fig:arrottPlots}}
\end{figure}
Figure \ref{fig:arrottPlots}(a) shows isothermal $M(H)$ curves of FeGe at selected temperatures for applied magnetic fields up to 50~kOe along the [100] direction.
The inherent curvature prevents a determination of the spontaneous magnetization $M_s(T)=M(T,H=0)$ from a linear extrapolation of the high-field data to $H=0$.
Also in conventional Arrott plots, i.e., presenting the data as $M^2$ vs.\,$H/M$, the isotherms are non-linear and concave (not shown).
Therefore they cannot be used to obtain $M_s(T)$ and $\chi^{-1}(T)$ from the intercepts of the ordinate and abscissa, respectively.
This implies that the magnetization of FeGe cannot be described by mean-field theory, i.e., with critical exponents $\beta=0.5$ and $\gamma=1$.
Thus, for the scaling analysis the isothermal magnetization data have to be re-scaled and presented as $M^{1/\beta}$ vs.\, $(H/M)^{1/\gamma}$ in a so-called modified Arrott plot.
For this purpose the critical exponents $\beta$ for the magnetization and $\gamma$ for the normalized field are chosen in such a way that the magnetization curves are linear \emph{and} parallel over a field range as large as possible with one set of critical exponents.
This is achieved in an iterative process.

As a first guess, exponents of the 3D-Heisenberg universality class were used to re-scale the isotherms.
Then a straight line was fitted to all re-scaled $M(H)$ curves.
In this fitting process magnetization data below a cut-off field $H_{\mathrm{\mathrm{cutoff}}}=2.5\times H_{c2}(T=0)=9$~kOe were not considered in order to minimize the influence of the chiral states on the high-field extrapolation and the determination of the critical exponents.
The intercepts of these linear fits with the ordinate and abscissa yield $M_s(T)^{1/\beta}$ and $\chi(T)^{-1/\gamma}$, respectively.
Their temperature dependence was analyzed according to
\begin{eqnarray}
M_s(T)       & = & M_s(0)\left(1-\frac{T}{T_c}\right)^{\beta}\, , \,T<T_c\, \label{eq:Mcritical}\\
\chi(T)^{-1} & = & \chi(0)^{-1} \left(\frac{T}{T_c}-1\right)^{\gamma}\, , \, T>T_c.  \label{eq:ChiCritical}
\end{eqnarray}
This gave new values for the exponents which were used in the next iteration step to re-scale the $M(H)$ data and to extract updated temperature dependencies of $M_s(T)$ and $\chi(T)^{-1}$.
This iteration process was repeated until the exponents between two iterations converged.
It is noteworthy that the iteration process converged to the same values of the exponents and $T_c$, independent of the initial values used if the fits to $M_s(T)$ and $\chi(T)^{-1}$ were constrained to the temperature interval $275~\mathrm{K}\leq T \leq 283~\mathrm{K}$.

Figure~\ref{fig:arrottPlots}(b) shows the re-scaled $M(H)$ isotherms using $\beta=0.368$ and $\gamma=1.382$ for temperatures $275~\mathrm{K}\leq T\leq 283~\mathrm{K}$.
In this narrow temperature range around $T_c$ ($\Delta T/T_c \lesssim 1\%$) the temperature increments between two isotherms with respect to $T_c$ is smaller than 0.1\%.
Only in these circumstances the slopes of the linearized magnetization curves are the same within 2\% and hence the isotherms are almost parallel lines.
Outside this temperature window however, the slopes steadily decrease and are up to 10\% smaller at 265~K and 300~K than at $T_c$.
These data were not considered as they lead to an overestimation of $M_s(T)$ and $\chi(T)^{-1}$.
As can be seen from Fig.~\ref{fig:arrottPlots}(b) the linearity of the data is perfectly adhered down to $H_{\mathrm{cutoff}}$ (large symbols).
However, an increasing deviation of the linear extrapolation from the data (small symbols) is seen upon approaching $H=0$ with increasing temperature.
This exemplifies the importance to constrain the fits to fields strong enough to overcome the DM and exchange interactions and to ensure that the magnetic moments are in a field-polarized state.
The curve measured at 278.6~K is extrapolating to the origin and hence it represents the critical isotherm (highlighted in red).
The positive slopes of the isotherms suggest that the magnetic ordering in FeGe is a continuous transition according to the Banerjee criterion\cite{Banerjee64}.
%

\begin{figure}
\includegraphics[width=8.5cm]{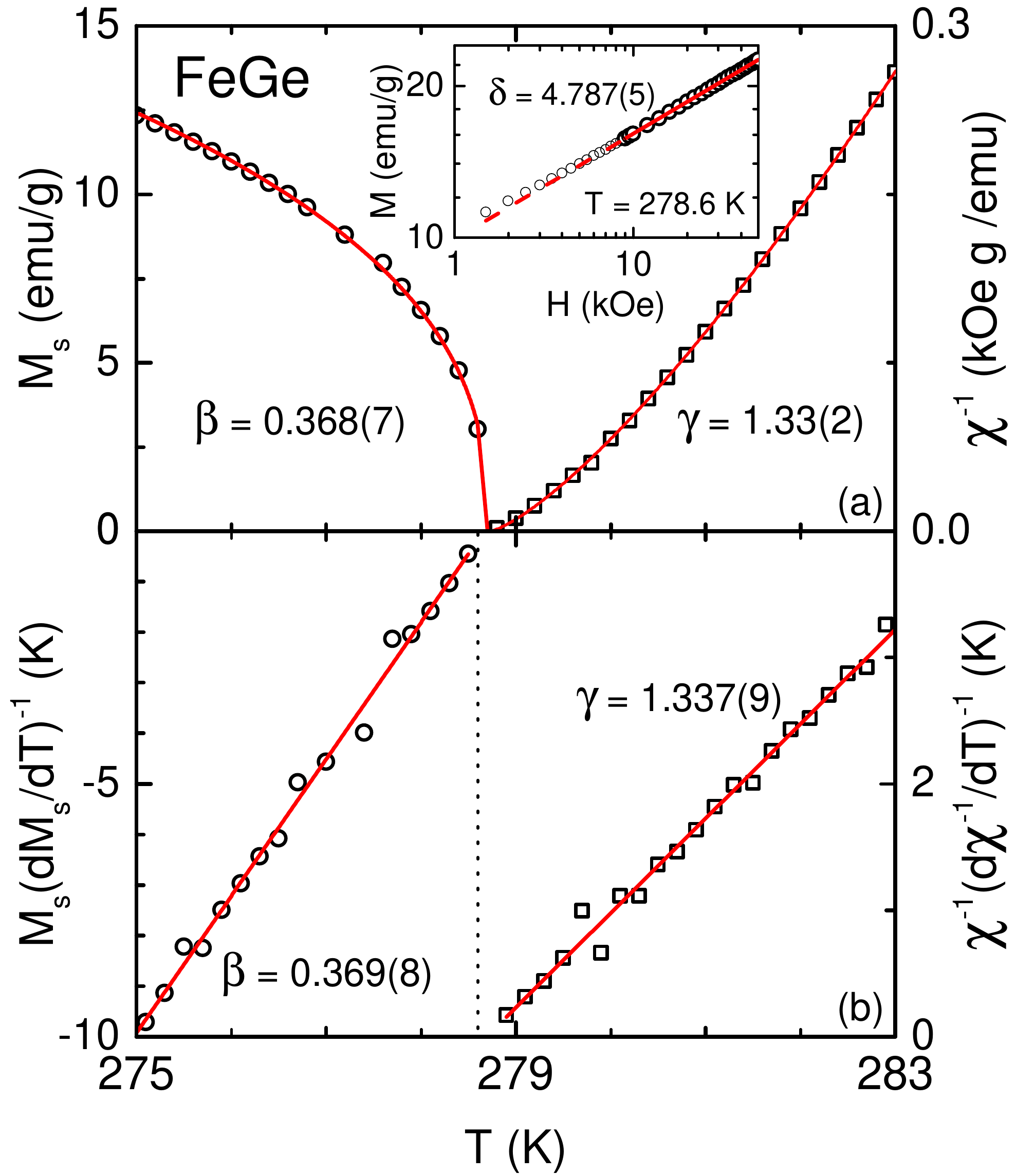}
\caption{(a) Temperature dependence of the spontaneous magnetization $M_s(T)$ (left) and inverse susceptibility $\chi(T)^{-1}$ (right) of FeGe.
The lines represent fits of  eq.~(\ref{eq:Mcritical}) and eq.~(\ref{eq:ChiCritical}) to the data with the exponents $\beta=0.368$ and $\gamma=1.382$.
The inset shows the critical isotherm $M(H)$.
Its slope yields the exponent $\delta=4.787$.
(b) The Kouvel-Fisher plots of $M_s(T)(dM/dT)^{-1}$ (left) and $\chi(T)^{-1}(d\chi(T)^{-1}/dT)^{-1}$ (right) yield the exponents $\beta=0.369$ and $\gamma=1.337$, respectively.
The vertical dashed line indicates $T_c=278.6$~K.
\label{fig:MandChi}}
\end{figure}

The temperature dependence of $M_s(T)$ and $\chi(T)^{-1}$ obtained from this re-scaling procedure is depicted in Fig.~\ref{fig:MandChi}(a).
The lines indicate fits of eq.~(\ref{eq:Mcritical}) and (\ref{eq:ChiCritical}) to these data with $\beta=0.368$ and $\gamma=1.33$.
Whereas $\beta$ is consistent with the value used to re-scale the isotherms, $\gamma$ is slightly smaller.
Nevertheless, both curves extrapolate to zero at $T_c=278.7$~K.
It is noted that the extrapolation of $\chi(T)^{-1}$ to higher temperature (not shown) starts to deviate from the data above 285~K.
However, the extrapolation of the $M_s(T)$ fit to lower temperature (not shown) yields a good description of the experimental data.

An alternative approach to extract the critical exponents and $T_c$ is to analyze the temperature dependencies of $M_s(dM_s/dT)^{-1}$ and $\chi^{-1}(d\chi^{-1}/dT)^{-1}$.
In these so-called Kouvel-Fisher plots\cite{Kouvel64,Fisher69} both quantities should depend linearly on temperature.
This is indeed the case and from the inverse of the slopes the critical exponents and from the intercept with the temperature axis $T_c$ can be inferred.
Fitting the data shown in Fig.~\ref{fig:MandChi}(b) yields $\beta=0.369(8)$ and $\gamma=1.337(9)$ in agreement with the values found for $M_s(T)$ and $\chi(T)^{-1}$ (see Fig.~\ref{fig:MandChi}(a)).

The critical exponent $\delta$ can be extracted from the critical isotherm as $M$ and $H$ are related by the Widom scaling hypothesis according to
\begin{equation}
M=DH^{1/\delta}\, , \label{eq:widom}
\end{equation}
with $D$ the critical amplitude\cite{Binney92,Stanley71,Campostrini02}.
This relation allows the critical exponent $\delta$ to be determined experimentally without using the scaling hypothesis that requires other critical exponents.
The inset to Fig.~\ref{fig:MandChi}(a) shows the critical isotherm in a double-logarithmic plot.
From a linear fit (solid line) to the data above $H_{\mathrm{cutoff}}=9$~kOe (large circles) $\delta=4.787(5)$ and $D=2.35(5)$ were obtained.
The extrapolation of the fit to lower fields (dashed line) starts to deviate from the data (small circles) as here the field is not strong enough to fully align the magnetic moments along the field direction.


\begin{figure}
\includegraphics[width=8.5cm]{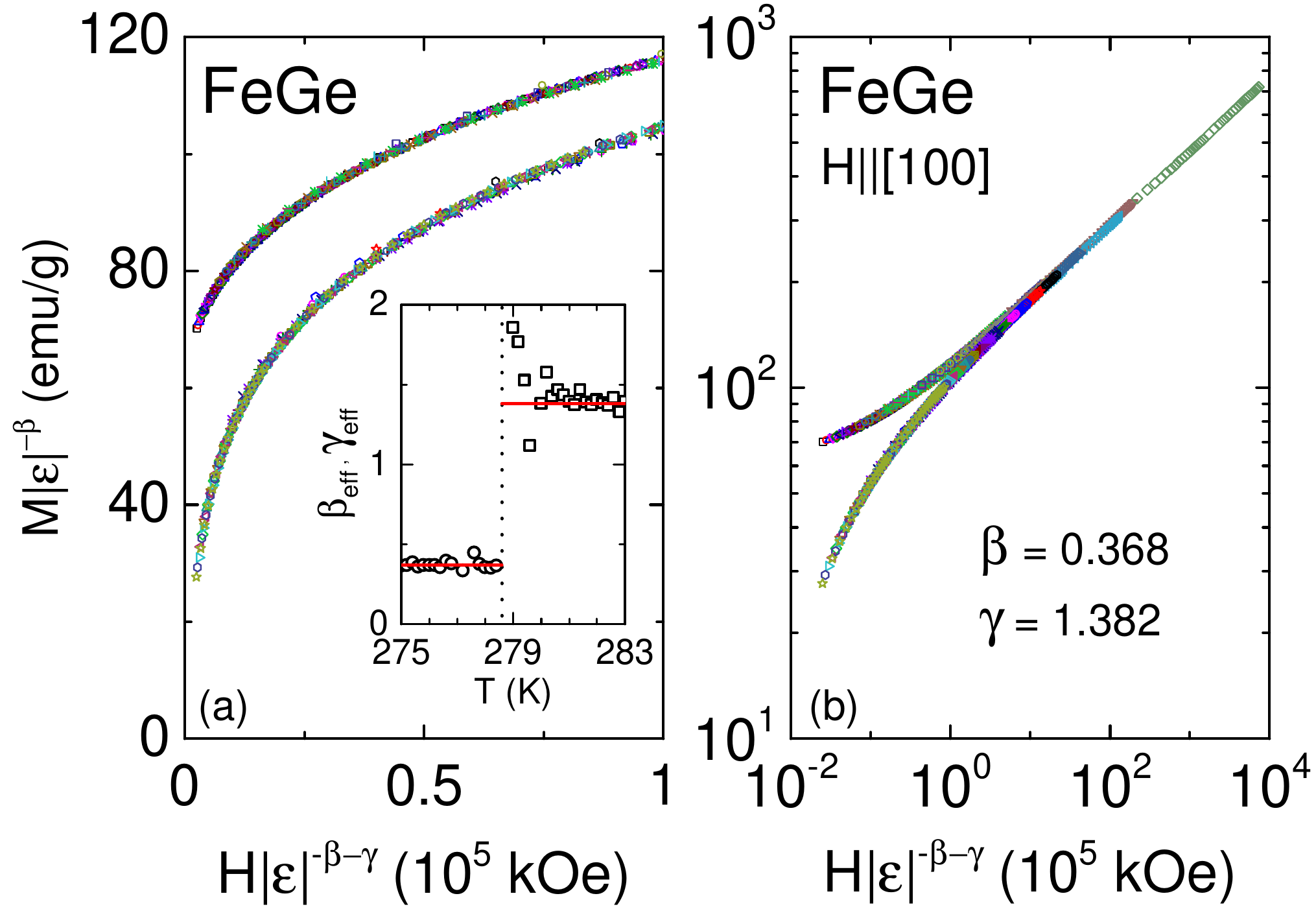}
\caption{Scaling plots of the $M(H)$ data of FeGe for temperatures $275~\mathrm{K}\leq T \leq 283~\mathrm{K}$ using the values $\beta=0.368$, $\gamma=1.382$ and $T_c=279.0$~K.
(a) All data fall on two universal curves.
The inset shows the effective exponents $\beta_{\mathrm{eff}}$ and $\gamma_{\mathrm{eff}}$ according to eq.~(\ref{eq:effectiveExponents}).
The vertical dashed line indicates $T_c$.
(b) Scaling plot in a double-logarithmic scale.
\label{fig:scalingPlots}}
\end{figure}

A stringent test to exemplify the quality of the scaling is a plot of the magnetic equation-of-state (EoS) in the critical region ($275~\mathrm{K}\leq T \leq 283~\mathrm{K}$).
The magnetic EoS is given by
\begin{equation}
M(H,T)=\left|1-\frac{T}{T_c}\right|^{\beta}\times f_{\pm}\left(\frac{H}{\left|1-\frac{T}{T_c}\right|^{\beta+\gamma}}\right)\,,
\label{eq_state}
\end{equation}
with regular analytic functions $f_+$ (for $T>T_c$) and $f_-$ (for $T<T_c$)\cite{Binney92,Stanley71,Campostrini02}.
Thus, if the scaling is consistent, all magnetization data fall on two universal curves.
Figure~\ref{fig:scalingPlots} shows this scaling where the $M(H)$ data are plotted as $M|\epsilon|^{-\beta}$ vs.\, $H|\epsilon|^{-\beta-\gamma}$.
It is obvious that the $M(T,H)$ data fall on two curves, depending on their temperature (Fig.~\ref{fig:scalingPlots}(a), top curve: $T<T_c$).
Extending the field range and representing the isotherms in a double-logarithmic plot reveals that all data collapse on a universal curve that bifurcates at $T_c$ (c.f. Fig.~\ref{fig:scalingPlots}(b)).

\begin{table}
\caption{The critical exponents $\beta$, $\gamma$, and $\delta$ as well as the critical temperature $T_c$ for FeGe obtained from the scaling analysis in a narrow temperature interval ($\Delta T/T_c <0.1\%$) around $T_c$.
From these exponents the values of $\delta$ and $\alpha$ were derived from $\beta$ and $\gamma$ via the scaling relations given in eq.~(\ref{eq:deltaScalingRelation}) and eq.~(\ref{eq:alphaScalingRelation}), respectively.
For comparison the exponents for several universality classes are also included.
\label{tab:exponents}}
\begin{tabular}{l|c|c|c|c|c}\hline \hline
method                         & $T_c$ (K) &$\beta$  &$\gamma$ & $\delta$  & $\alpha$  \\ \hline
Fig.~1(b)                      & 278.6     & 0.368    & 1.382  &  -        &   -       \\
eq.~(\ref{eq:deltaScalingRelation},\ref{eq:alphaScalingRelation})&-&-&-& 4.755 & -0.12 \\ \hline
$M_s(T)$                       & 278.7(3)  & 0.368(7) & -      & -         & -         \\
$\chi(T)^{-1}$                 & 278.7(1)  & -        & 1.33(2)& -         & -         \\
eq.~(\ref{eq:deltaScalingRelation},\ref{eq:alphaScalingRelation})&-&-&-& 4.6(1) & -0.07(3)\\ \hline
$M(dT/dM)$                     & 278.7     & 0.369(8) & -       &   -      & -         \\
$\chi^{-1}(dT/d\chi^{-1})$     & 278.7     & -        &1.337(9) &  -       & -          \\
eq.~(\ref{eq:deltaScalingRelation},\ref{eq:alphaScalingRelation})&-&-&-& 4.62(8) & -0.08(2) \\ \hline
$M=DH^{1/\delta}$               & 278.6(1) &   -     &   -     & 4.787(5) & -          \\ \hline
eq.~(\ref{eq_state})            & 279.0(1) & 0.368   & 1.382   & -        & -          \\ \hline
3D-Heisenberg\cite{Campostrini02}& -      & 0.3689  & 1.396   & 4.783    & -0.133      \\
3D-Ising\cite{Guida97}         & -        & 0.3250  & 1.241   & 4.817    & 0.109       \\
3D-XY Ising\cite{Leguillou77}  & -        & 0.3454  & 1.316   & 4.810    & -0.01       \\ \hline \hline
\end{tabular}
\end{table}

To rule out any significant influence of competing magnetic couplings and/or randomness on the critical behavior, the effective exponents\cite{Stanley71}
\begin{equation}
\beta_{\mathrm{eff}}=\frac{d(\ln M_s(\epsilon))}{d(\ln
\epsilon)},\,\gamma_{\mathrm{eff}}=\frac{d(\ln \chi(\epsilon)}{d(\ln \epsilon)}
\label{eq:effectiveExponents}
\end{equation}
were determined as well.
They are inferred from the generalized power laws for the critical behavior and approach universal critical exponents in the limit $\epsilon\rightarrow 0$.
The effective exponents shown in the inset to Fig.~\ref{fig:scalingPlots}(a) are consistent with the values of $\beta$ and $\gamma$ (indicated by the lines) obtained above and rule out any competing magnetic interaction.

The various critical exponents and temperatures determined above are summarized in Tab.~\ref{tab:exponents}.
In addition, the critical scaling hypothesis\cite{Binney92}
\begin{eqnarray}
\delta &=& 1+\frac{\gamma}{\beta}
\label{eq:deltaScalingRelation}
\end{eqnarray}
allows $\delta$ to be calculated for each set of $\beta$ and $\gamma$ exponents derived experimentally.
Overall, a consistent set of exponents is found.
However, the $\gamma$ values obtained from the Kouvel-Fisher plot (Fig.~\ref{fig:MandChi}) are slightly smaller as those used in the re-scaled $M(H)$-curves (Fig.~\ref{fig:arrottPlots}(b)) and the scaling plot (Fig.~\ref{fig:scalingPlots}).
As a consequence, the values of $\delta$ are also slightly smaller than the one inferred from the critical isotherm.
But eq.(\ref{eq:deltaScalingRelation}) can be used to calculate $\gamma$ by using the consistent values $\beta=0.368$ (from Fig.~\ref{fig:arrottPlots}(b) and Fig.~\ref{fig:scalingPlots}) and $\delta=4.787$ from the isotherm (inset Fig.~\ref{fig:MandChi}(a)).
This yields $\gamma=1.39(3)$.
Thus, this provides convincing evidence that the set of critical exponents for FeGe inferred from the $M(H)$ data is very close to the values predicted for the 3D-Heisenberg universality class\cite{Campostrini02}.
A clear deviation from the exponents of the other universality classes is obvious.
Therefore, it can be concluded that FeGe is an isotropic 3D-Heisenberg ferromagnet and does not change universality class across the transition as reported recently\cite{Zhang16}.
Furthermore, this scaling study confirms that (i) the magnetic system of these chiral magnets is basically that of a simple ferromagnet and (ii) in the hierarchy of the magnetic couplings the DM interaction is much weaker than the isotropic exchange.

\begin{figure}
\includegraphics[width=8.5cm]{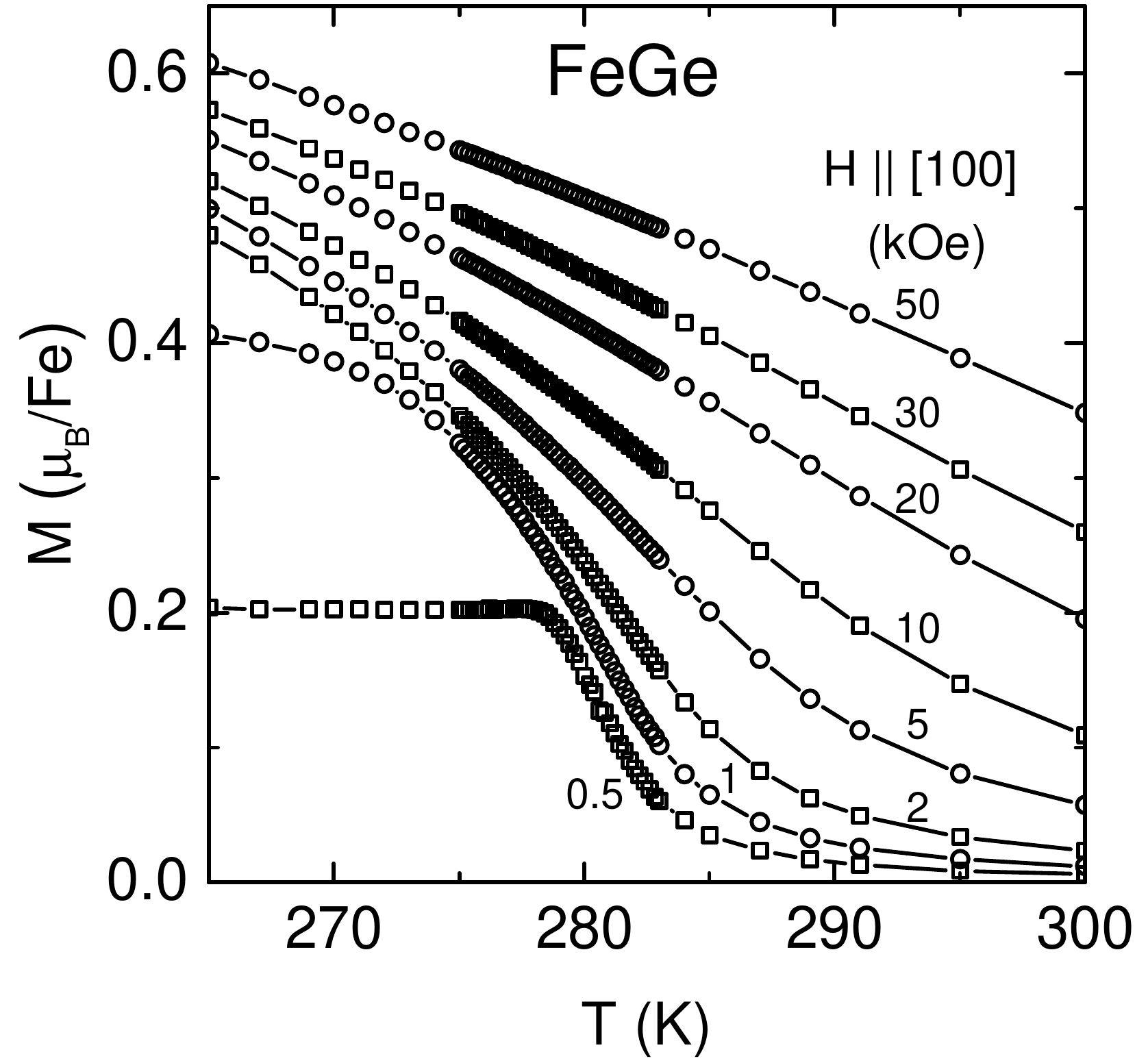}
\caption{Temperature dependence of the magnetic moment of FeGe at various magnetic fields.
These data were obtained from the $M$ vs $H$ data shown in Fig.~\ref{fig:arrottPlots}(a).
\label{fig:MvsTanddMdT}}
\end{figure}

The different methods to analysis the $M(H)$ data led to a critical temperature $T_c=278.8(2)$~K (Tab.~\ref{tab:exponents}) that is slightly higher than $T_c=278.2(3)$ obtained from ac-susceptibility measured in zero field\cite{Wilhelm11}.
This is not surprising as the relatively strong magnetic fields used to induce a collinear spin alignment will alter the correspondent PM to helical-ordering transition temperature.
The latter has to be different to the one found for the crossover from the ICS state to the helical phase in zero field.

This study shows that FeGe belongs to the 3D-Heisenberg universality class like the other chiral magnets Fe$_{0.8}$Co$_{0.2}$Si\cite{Jiang10} and Cu$_2$OSeO$_3$\cite{Zivkovic14}.
However, in contrast to this, MnSi has been argued to exhibit tri-critical mean-field behavior\cite{Zhang15}.
Thus, the question arises why MnSi should be so different from the other cubic helimagnets despite the same underlying magnetic interactions.
Unfortunately, this question cannot be unambiguously answered as no scaling study has been made in the appropriate regime for MnSi so far.
As pointed out above, a scaling analysis can only provide a meaningful answer if it is done in the critical region ($\epsilon \approx 10^{-2}$) and on the unperturbed magnetic system, i.e. for fields well above $H_{c2}$.
These requirements are obviously quite challenging for MnSi, given its low $T_c$ and relatively high $H_{c2}$ values, and were not fulfilled in the reported scaling studies\cite{Zhang15,Chattopadhyay09}.
Moreover, the spin polarization of the weak itinerant band-ferromagnet MnSi\cite{Shinoda66} is strongly affected by the longitudinal spin-fluctuations in contrast to the \emph{stable} spin-polarization of Fe in the strong band-ferromagnet FeGe\cite{Lundgren70,Ericsson81,Barla15,Yamada03,Neef09}.
This effect is seen in a very large high-field susceptibility in MnSi which further complicates the scaling analysis.
Thus, a detailed and careful investigation close to $T_c$ is required to unveil the critical behavior of the putative PM-to-FM transition in MnSi.

The $M(H)$ isotherms of FeGe can also be used to determine the temperature variation of the magnetization as shown in Fig.~\ref{fig:MvsTanddMdT}.
The $M(T)$ curve measured in $H=0.5$~kOe, the lowest field in this experiment, shows a steep increase close to $T_c$, identical to literature data\cite{Cevey13}.
At this field and below $T_c$, FeGe enters the conical phase and the magnetization becomes constant\cite{Wilhelm11,Wilhelm12,Cevey13}.
At 1~kOe, the temperature range of the present study was not extending to low enough temperatures to reach this plateau.
Fields of 5~kOe and above are well beyond $H_{c2}$, where FeGe will be always in the field-polarized state\cite{Wilhelm11}.
These data will be analyzed and discussed further below.

\subsection{Specific Heat and Thermal Expansion}

\begin{figure}
\hspace{-5mm}
\includegraphics[width=0.475\textwidth]{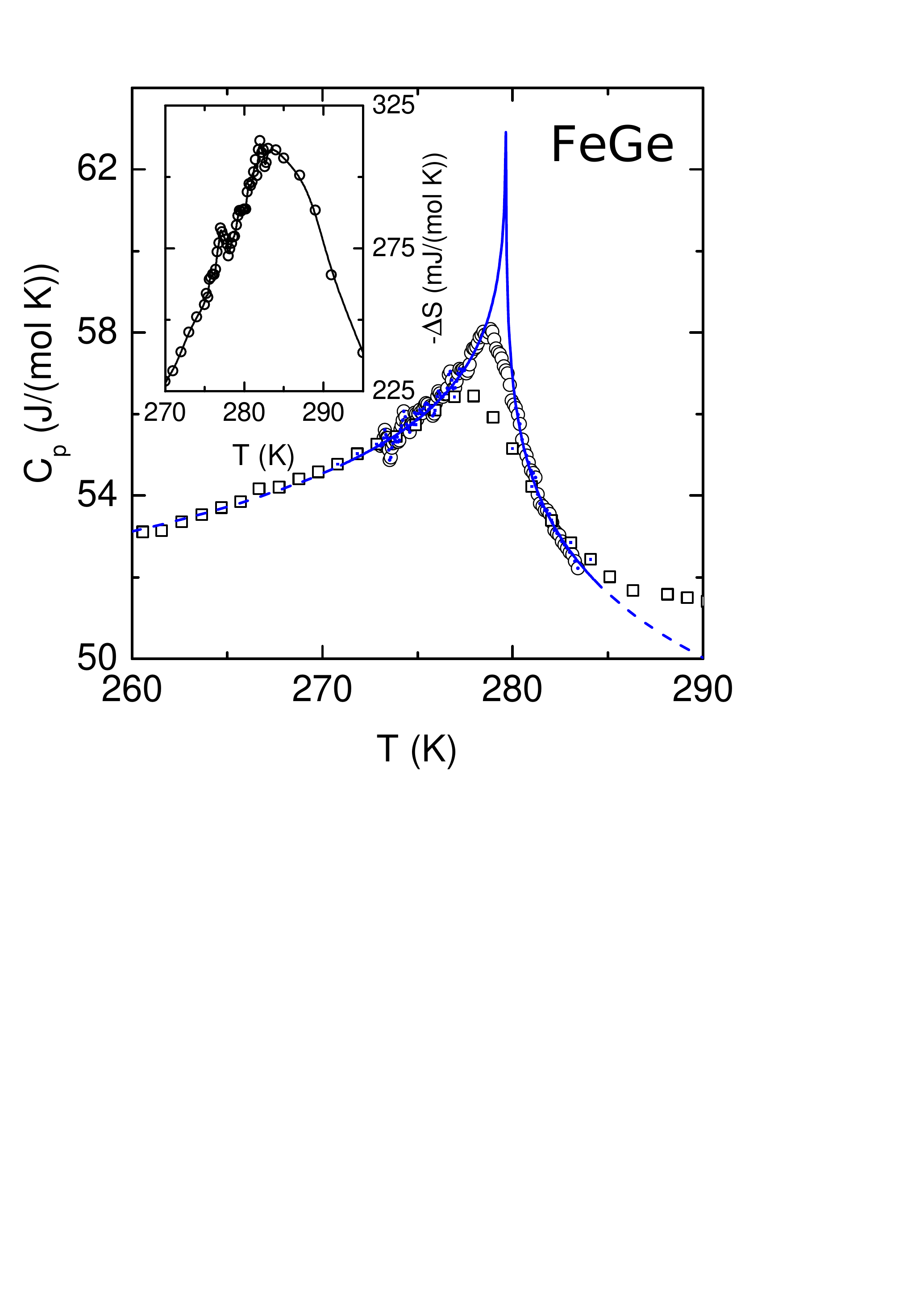}
\caption{
Specific heat $C_p(T)$ of FeGe in the vicinity of $T_c$.
Two data sets show a cusp at 278.0~K (squares) and 278.7~K (circles).
The solid line represents a fit of eq.(\ref{eq:CpCritical}) with $\alpha=-0.133$ and $T_c=279.7$~K to those data points marked with a dot; its extrapolation is shown as dashed line.
The inset shows the change of the entropy ($-\Delta S(T,\Delta H$)) derived from the isothermal magnetization data via eq.~(\ref{eq:entropy}) (see text).
\label{fig:critical}}
\end{figure}

The specific heat data shown in Fig.~(\ref{fig:critical}) were also included in the scaling analysis.
The  $C_p(T)$ data near $T_c$ can be analyzed according to
\begin{equation}
C_{\mathrm{mag}}^{\mathrm{crit}}(T)  = C_0+\left\{
\begin{array}{ll}
A_+\left(\frac{T}{T_c}-1\right)^{-\alpha}, & T>T_c \\
A_-\left(1-\frac{T}{T_c}\right)^{-\alpha}, & T<T_c \\
\end{array}
\right.\label{eq:CpCritical}
\end{equation}
with the critical exponent $\alpha$, the critical amplitudes $A_\pm$\cite{Binney92,Stanley71}, and a constant background $C_0$.
For this analysis two data sets measured in zero field were used.
In order to eliminate precursor effects close to $T_\mathrm{c}$\cite{Wilhelm11,Wilhelm12}, data points in the range 277.15~K$<T<280.15$~K have been excluded.
A reasonable fit of eq.~(\ref{eq:CpCritical}) to the data was possible with a fixed value $\alpha=-0.133$, i.~e.,~the critical exponent for isotropic 3D-Heisenberg magnets\cite{Campostrini02} (solid line in Fig.~\ref{fig:critical}).
The best fit was obtained using $T_c=279.7$~K, $A_+=-29.7$~J/mol/K, $A_-=-22.4$~J/mol/K, and $C_0=68.9$~J/(mol K).
This fit results in a $\lambda$-like peak that is located slightly above the temperature where the experimental $C_p(T)$ data attain a maximum.
The deviation from the experimental data points very close to the maximum is expected as the fit describes a PM-to-FM phase transition whereas the helical fluctuations in the ICS phase are present in the experimental data.
The ratio of the amplitudes $A_+/A_-=1.3$ is close to ratios known from other experiments ($1.40\leq A_+/A_-\leq 1.52$)\cite{Kaul91} and theory ($A_+/A_-=1.56$) for the 3D-Heisenberg universality class\cite{Campostrini02}.

The exponent $\alpha$ describing the critical behavior of the magnetic part of the specific heat is connected to exponents obtained from the scaling analysis in Sec.~\ref{sec:arrott} via the scaling relation\cite{Binney92,Stanley71,Campostrini02}
\begin{equation}
\alpha = 2-2\beta-\gamma\,. \label{eq:alphaScalingRelation}\\
\end{equation}
With this relation $\alpha$ can be calculated from the sets of exponents given in Tab.~\ref{tab:exponents}.
This gives $\alpha\approx -0.08(2)$ for the exponents obtained from the modified Arrott and Kouvel-Fisher plots (see Tab.~\ref{tab:exponents}).
Although this value is only in fair agreement with the value used to fit the specific heat, it seems to be obvious that values for $\alpha$ corresponding to other universality classes are not able to describe the specific heat data of FeGe.

\begin{figure}
\includegraphics[width=0.475\textwidth]{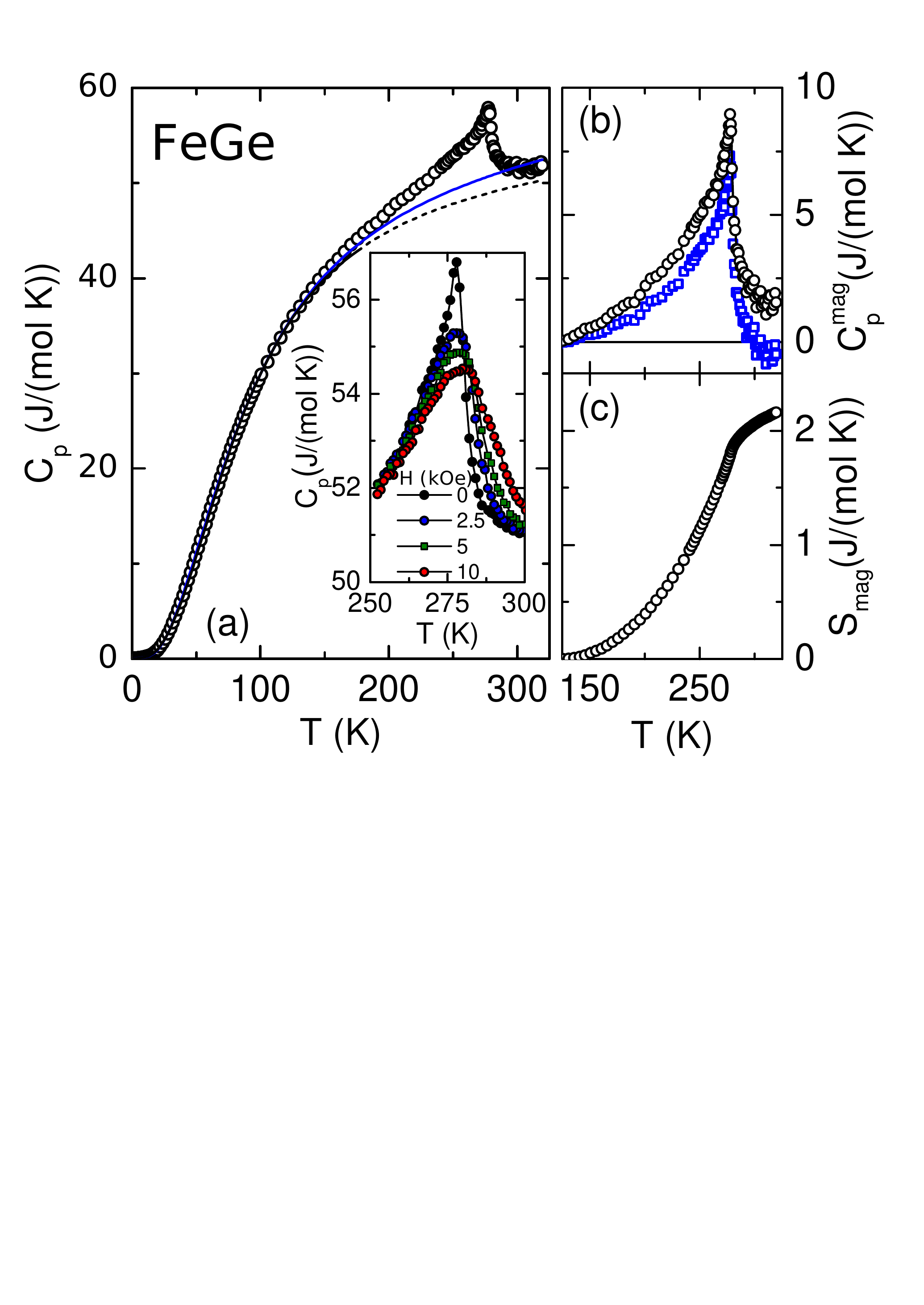}
\caption{(a) The specific heat $C_p(T)$ of FeGe (circles).
Non-magnetic contributions to the specific heat are well described by a Einstein-Debye model (dashed line) and
by \emph{ab initio} phonon DoS calculations (solid line).
Inset: $C_p(T)$ in the vicinity of $T_c$ at various magnetic fields.
(b) Magnetic contribution to the specific heat, $C_p^{\mathrm{mag}}(T)$, after subtraction of the electronic part and lattice contributions, based on a Debye-Einstein model (circles) or the phonon DOS calculation (squares).
(c) Magnetic entropy $S_{\mathrm{mag}}$ determined from the integration $C_p^{\mathrm{mag}}(T)/T$ (circles in (b)).
\label{fig:cp}}
\end{figure}

The $M(T)$ data shown in Fig.~\ref{fig:MvsTanddMdT} can be used to determine the change in entropy $\Delta S(T)$ across the phase transition.
One of the fundamental Maxwell relations of thermodynamics relates the field-induced entropy change at constant temperature with the temperature derivative of the magnetization at constant field.
This leads to
\begin{equation}\label{eq:entropy}
\Delta S(T, \Delta H)= \int_0^H{\left(\frac{dM}{dT}\right)_{H'} dH'}\,.
\end{equation}
Thus, this is an indirect method to calculate the magneto-caloric effect (MCE) and to measure the decrease of the total entropy during an isothermal field sweep.
$\Delta S(T,\Delta H)$ was determined according to eq.~(\ref{eq:entropy}) by numerical integration of the data shown in Fig.~\ref{fig:MvsTanddMdT} and is depicted in the inset to Fig.~\ref{fig:critical}.
It shows a broad maximum centered at 283~K.
This indicates, that the transition from the PM into the magnetically ordered state in zero field is broadened by the helical fluctuations prevailing in the ICS phase.
Compared to other $B$20 helimagnets, like Fe$_{1-x}$Co$_x$Si alloys\cite{Chattopadhyay02}, the MCE is about three times as large but in contrast to rare-earth based ferromagnets it is considerably smaller\cite{Pecharsky99}.

The $C_p(T)$ data of FeGe down to 4.2~K presented in Fig.~\ref{fig:cp}(a) allow further details to be extracted and to be compared with quasi-harmonic \emph{ab initio} phonon calculations.
The transition into the magnetically ordered phase is seen as pronounced cusp centered at $T_c=278.0(5)$~K on top of the lattice contribution to the specific heat.
Above the transition, at 298~K the specific heat, entropy, and enthalpy are $C_p=51.7(10)$~J/mol/K, $S=65.7(11)$~J/mol/K, and $H=10.27(20)$~J/mol, respectively.
The anomaly caused by the transition broadens considerably in an external field and seems to shift steadily towards higher temperatures (280~K at 10~kOe, see inset to Fig.~\ref{fig:cp}(a)).

The low-temperature part of the specific heat in zero field is decomposed into an electronic $C_{\mathrm{el}}(T) = \gamma_{\mathrm{el}}T$ and a lattice contribution $C_p^{\mathrm{lat}}(T)= \beta_{\mathrm{lat}}T^3$, $C_p(T)=C_p^{\mathrm{el}} +C_p^{\mathrm{lat}}$.
From a plot of $C_p/T$ vs. $T^2$ (not shown) the Sommerfeld coefficient $\gamma_{\mathrm{el}}=10.6$ mJ/(mol K$^2$) and the value $\beta_{\mathrm{lat}}=6.56\times10^{-5}$ J/(mol K$^4$) were obtained.
They are in a good agreement with literature data\cite{Marklund74,Yeo03,DiTusa14}.
The small value of $\gamma_{\mathrm{el}}$ together with other experimental\cite{Lundgren70,Ericsson81,Barla15} and theoretical\cite{Yamada03,Neef09} findings manifests the stable character of the Fe moments like in a strong band-ferromagnet.
From $\beta_{\mathrm{lat}}$ the Debye temperature is calculated to $\Theta_D=390$~K.
Fitting a Debye function to the data up to 175~K with a fixed $\gamma_{\mathrm{el}}$ yields $\Theta_D=348$~K, also in agreement with reported values\cite{Marklund74,Yeo03}.
A weighted sum of Debye function and Einstein mode in the same temperature interval gives $\Theta_D=370$~K and $\Theta_E=137$~K with $w=0.1$, the weighted contribution of the Einstein mode (dashed line in Fig.~\ref{fig:cp}(a).
This describes the data quite well but it underestimates the heat capacity above $T_c$ slightly.

\begin{widetext}
\begin{figure*}
\includegraphics[width=1\textwidth]{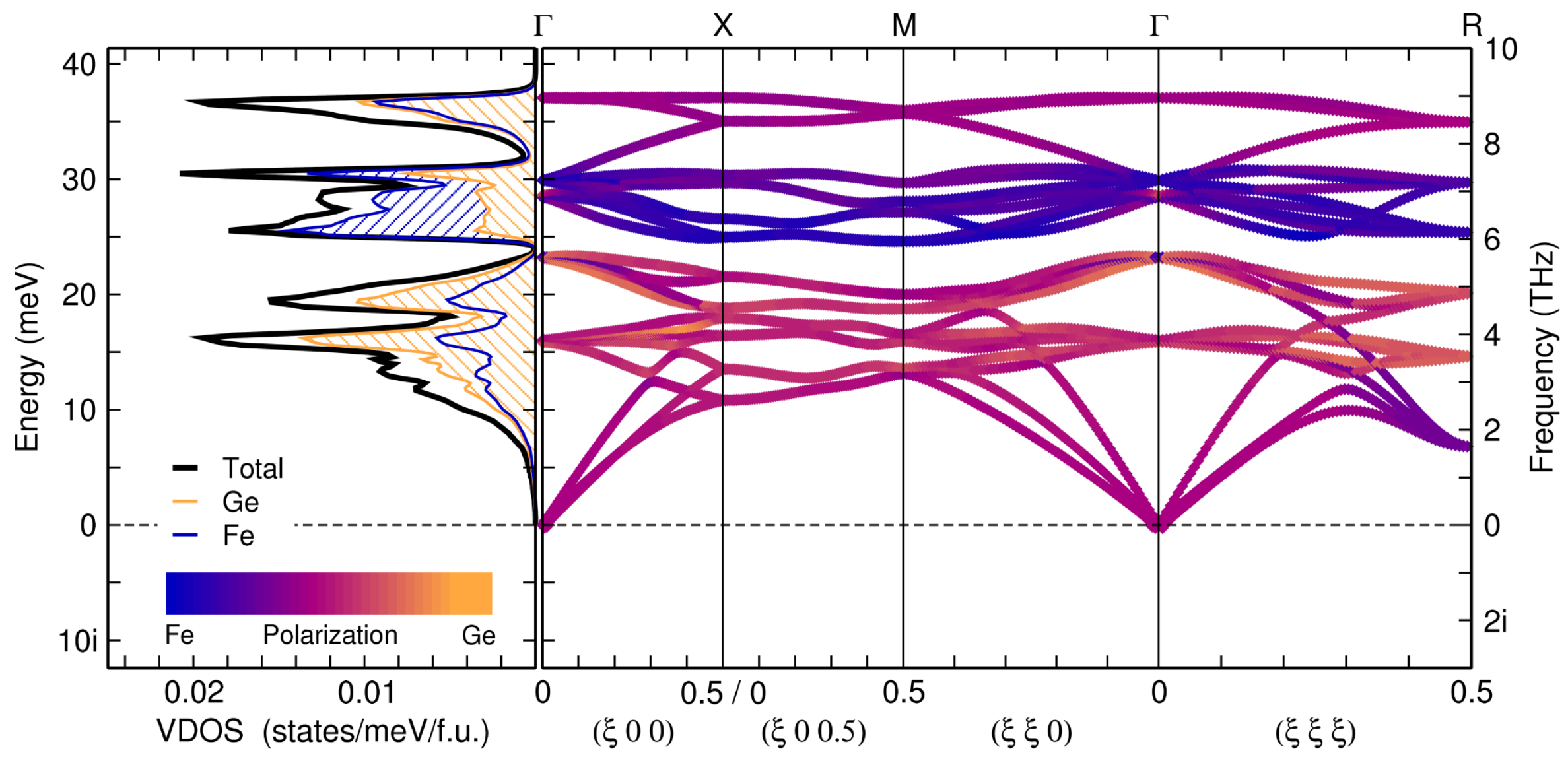}
\caption{Calculated vibrational density of states (VDOS) and phonon dispersion of FeGe along lines between selected high symmetry points in the Brillouin zone.
The color coding refers to the elemental character of the phonon modes (blue for Fe and orange for Ge).
\label{fig:phonon}}
\end{figure*}
\end{widetext}

In order to have a better estimate for the phonon contribution to the specific heat, lattice vibrations were calculated from first-principles.
For the phonon dispersion and vibrational density of states shown in Fig.~\ref{fig:phonon}, optimized structural parameters, i.e., an equilibrium lattice constant of $a=4.669$\,\AA, internal structural parameters of the \emph{B}20 structure (space group P2$_1$3) $u=0.1354$ for Fe, $u=0.8418$ for Ge, and a bulk modulus $B_0=158$~GPa were used.
These values as well as the electronic density of states (not shown) agree excellently with experiment\cite{Wilhelm07,Pedrazzini07} and previous DFT results\cite{Neef09}.
The configuration is dynamically stable and imaginary frequencies are absent.
According to the difference in the masses, the energies below 24 meV are dominated by Ge modes, while Fe modes dominate at higher frequencies.

Adding the experimentally determined electronic specific heat to the calculated phonon contribution yielded the non-magnetic contribution to $C_p(T)$ shown as solid line in Fig.~(\ref{fig:cp}(a)).
It is very close to the experimental data, also above the magnetic transition.
From this it becomes apparent that the magnetic contribution to the specific heat, $C_p^{\mathrm{mag}}(T)=C_p(T)-C_p^{\mathrm{lat}}(T)-C_p^{\mathrm{el}}(T)$, extends over a remarkably wide temperature range down $T\approx 150$~K (Fig.~\ref{fig:cp}(b)).
Thus, apart from the transition at $T_c$ the complex re-orientation of the magnetic structure, commencing below  about 245~K\cite{Lebech89}, seems to contribute and broaden $C_p^{\mathrm{mag}}(T)$ considerably.
This is also highlighted by $S_{\mathrm{mag}}(T)$, the magnetic part of the total entropy, plotted in Fig. \ref{fig:cp}(c).
At $T_c$ it contributes about 3\% to the total entropy.

From the electronic contribution to the specific heat, $\gamma_{\mathrm{el}}$, the electronic DoS at the Fermi level in a free-electron model is obtained as $N(E_\mathrm{F})=17.8$~states/(eV unit cell).
This value should be compared to $N(E_\mathrm{F})=9.6$~states/(eV unit cell) obtained from spin-polarized LDA band calculations\cite{Jarlborg04}.
Taking a reasonable estimate for the electron-phonon coupling $\lambda=0.2$ into account\cite{Jarlborg04} results in $\gamma_{\mathrm{el}}=\frac{1}{3}\pi^2k_B^2N(E_\mathrm{F})(1+\lambda)\approx 7$~mJ/(mol K$^2$).
This compares quite well with the experimental value and the rather small value of $\gamma_{el}$ shows that spin fluctuations play a minor role in the electronic properties of FeGe.
The electron mass enhancement of FeGe can be estimated to $m^*/m_0\approx 10$, using the calculated value for $\gamma_0=1$~mJ/(mol K$^2$) under the assumption of two free electrons per FeGe.

\begin{figure}
\hspace{-6mm}
\includegraphics[width=0.45\textwidth]{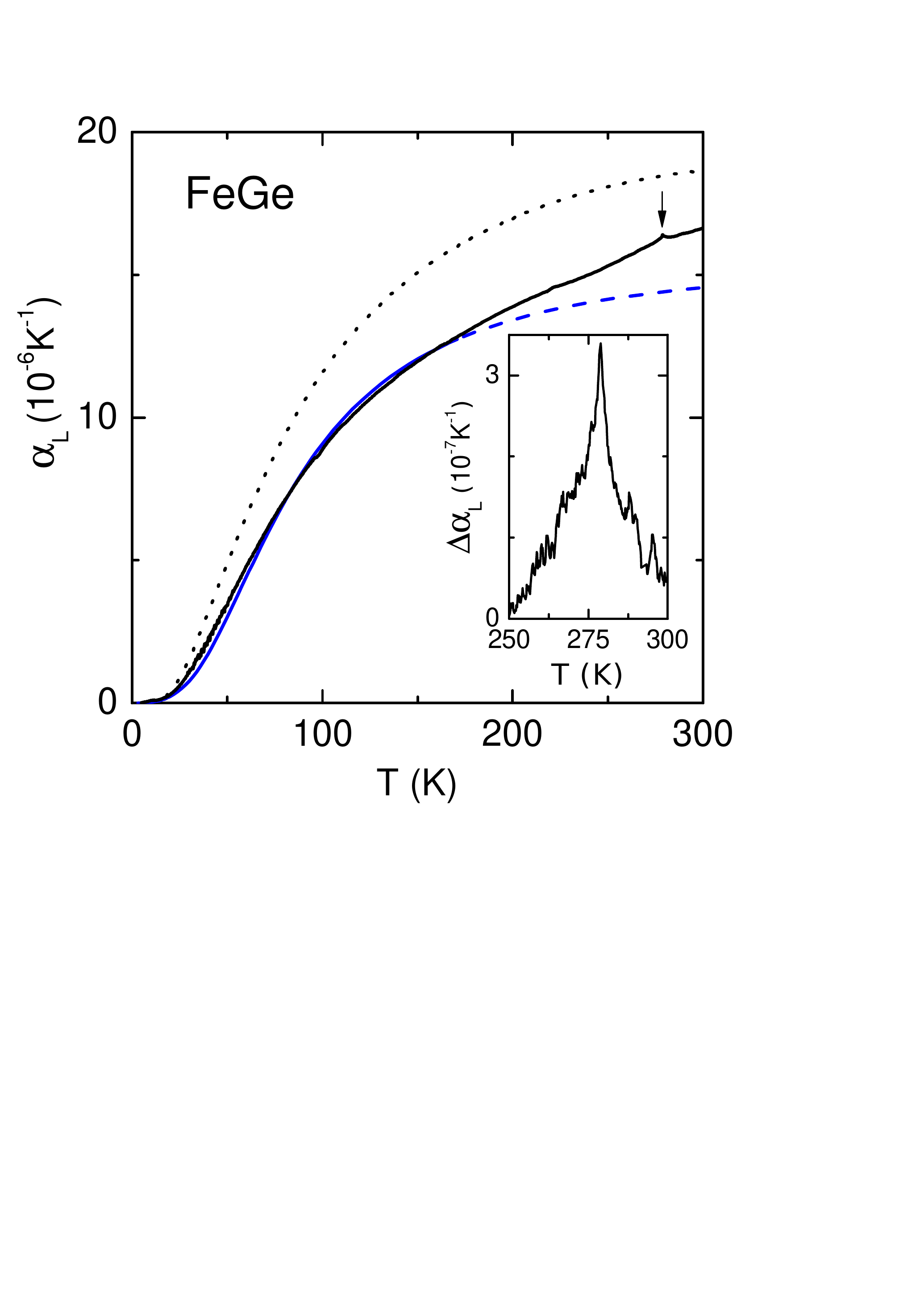}
\caption{Linear thermal expansion $\alpha_L(T)$ of FeGe.
The experimental data (line) show a small cusp at 278.8~K (arrow) that is caused by the magnetic transition.
The lattice contribution (dashed line) to the thermal expansion was obtained from the specific heat data according to eq.(\ref{eq:grueneisen})).
The dotted line is based on the phonon DOS calculations and represents $\alpha_L(T)=\alpha_V(T)/3$.
Inset: Experimental $\Delta\alpha_L(T)$ data close to $T_c$ after the subtracting of a linear
background from $\alpha_L(T)$.
\label{fig:thermExpanAndSXRD}}
\end{figure}

Figure~\ref{fig:thermExpanAndSXRD} shows the temperature dependence of the coefficient of the linear thermal-expansion $\alpha_L=(1/L)(dL/dT)$.
A small cusp in $\alpha_L(T)$ develops at the magnetic transition ($T=278.8$~K) as depicted in the inset to Fig.~(\ref{fig:thermExpanAndSXRD}).
The 50\% criterion was used to approximate the temperature where the jump occurred ($T=281.3$~K).
From the extrapolated anomaly it is obvious that its height is positive.
Together with the positive contribution $\Delta(C/T)=22$~mJ/mol/K$^2$ at the anomaly in the specific heat
if follows from the Ehrenfest relation, $dT_c/dp \propto \Delta\alpha_L/\Delta(C/T)$, that $T_c$ will initially increase with pressure.
A pressure study on FeGe revealed that $T_c$ was already suppressed to 274~K at 2.1~GPa\cite{Pedrazzini07}, i.e., well below its ambient pressure value.
Thus, the thermal expansion data imply that $dT_c/dp$ will change its sign and becomes negative within about 2~GPa.

The lattice contribution to the linear thermal expansion can be calculated using
$\alpha_L(T)= \alpha_V(T)/3$ with
\begin{equation}\label{eq:grueneisen}
\alpha_V(T) =\Gamma\kappa C_p^{\mathrm{lat}}(T)/V_m
\end{equation}
where $C_p^{\mathrm{lat}}(T)$ is the experimentally determined lattice contribution to the specific heat, $\Gamma$ is the phonon-Gr\"{u}neisen parameter, and $\kappa=-(1/V)dV/dp$ the isothermal compressibility.
Using the experimentally determined value $\kappa^{-1}=130~\mathrm{GPa}$\cite{Wilhelm07} and setting $\Gamma=1.9$ results in the $\alpha_L(T)$ shown in Fig.~\ref{fig:thermExpanAndSXRD} (dashed line).
This is in very good agreement with the experimental $\alpha_L(T)$ behavior for $T\lesssim 175$~K, i.e.~below the temperature where the helical propagation direction flips ($T\approx 210$~K)\cite{Lebech89}.
Thus, magneto-volume effects, caused by establishing the magnetic order, give an additional contribution to $\alpha_L(T)$ at higher temperatures.
The thermal expansion estimated from first-principles (dotted line in Fig.~\ref{fig:thermExpanAndSXRD}) overestimates the experimental data by about 12\% at $T_c$, which can still be considered an acceptable agreement.
Deviations must be expected from finite temperature changes to the electronic structure and the magnetic configuration or phonon-phonon interactions, which were not taken into account in this quasi-harmonic approach.

\section{Conclusion}

Isothermal magnetization data of FeGe in close vicinity of the helical ordering transition have been used for a thorough scaling analysis.
The re-scaled $M_s(T)^{1/\beta}$ vs. $(H/M)^{1/\gamma}$ curves yield straight and almost parallel lines for $\beta=0.368$ and $\gamma=1.382$ and result in scaling plots with universal curves.
The exponent $\delta=4.785$ was inferred from the critical isotherm.
Furthermore, similar values of exponents were obtained from the temperature dependence of $M_s(T)$ and $\chi^{-1}(T)$ as well as from corresponding Kouvel-Fisher plots.
Thus, this self-consistent scaling analysis manifests that the magnetic ordering in FeGe falls in the isotropic 3D-Heisenberg universality class.
The exponent $\alpha=-0.133$ of this universality class describes very well the specific heat anomaly in the vicinity of $T_c$.
The overall temperature dependence of the lattice contribution to the specific heat is accounted for by \emph{ab initio} phonon calculations.
It becomes apparent that the magnetic contribution to the specific heat is noticeable above 150~K which
reflects the complex reorientation of the helical propagation vector taking place in this temperature region.
Based on these results a quantitative phenomenological model of the magnetic structures of FeGe can now be established.

\begin{acknowledgments}
We acknowledge fruitful discussions with Yu. Grin and technical assistance from
S. Hoffmann, R. Koban, Y. Prots, H. Rave, and S. Scharsach.
Support by DFG project RO 2238/9-1 and GR 3498/3 (SPP1599) is gratefully acknowledged.
Calculations were carried out on the Cray XT6/m supercomputer of the Center for Computational Sciences and Simulation (CCSS) at the University of Duisburg-Essen.
\end{acknowledgments}


\begin{thebibliography}{99}
\bibitem{Tsteps} The $M(H)$ curves in panel (a) are separated by 2~K up to 277~K, by 0.8~K up to 281~K, by 1~K up to 285~K and by 2~K up to 291~K. The second last isotherm is measured at 295~K. The isotherms in (b) are separated by 0.8~K.
\end{thebibliography}
\end{document}